\documentclass[3p,times,twocolumn]{elsarticle}
 \biboptions{comma,sort&compress}
 
\usepackage{graphicx}
\usepackage{amsmath}
\usepackage{here}
\usepackage{ecrc}
\allowdisplaybreaks

\volume{00}

\firstpage{1}

\journalname{Nuclear and Particle Physics Proceedings}
\runauth{Nils Hermansson--Truedsson}

\jid{nppp}
\jnltitlelogo{Nuclear and Particle Physics Proceedings}

\DeclareUnicodeCharacter{2212}{\textendash}

\usepackage{amssymb}

\usepackage[figuresright]{rotating}
\begin{document}

\begin{frontmatter}

\title{
%
2-loop short-distance constraints for the $g-2$ HLbL \,$^*$} 
 
 \cortext[cor0]{Talk presented at QCD21 - 24th International Conference in QCD (5-9/07/2021,
  Montpellier - FR). }

 \author[label2]{Johan Bijnens}
 \address[label2]{Department of Astronomy and Theoretical Physics, Lund University, S\"{o}lvegatan 14 A, 223 62, Lund, Sweden}
\ead{bijnens@thep.lu.se}

 \author[label1]{Nils Hermansson--Truedsson\corref{cor1}}
\address[label1]{Albert Einstein Center for Fundamental Physics, Institute for Theoretical Physics, Universit\"{a}t Bern, Sidlerstrasse 5, 3012 Bern, Switzerland}
\cortext[cor1]{Speaker, Corresponding author}
\ead{nils@itp.unibe.ch}

\author[label1]{Laetitia Laub}
\ead{laub@itp.unibe.ch}

\author[label3]{Antonio Rodr\'{i}guez--S\'{a}nchez}
\address[label3]{Universit\'{e} Paris-Saclay, CNRS/IN2P3, IJCLab, 91405 Orsay, France}
\ead{arodriguez@ijclab.in2p3.fr}

\pagestyle{myheadings}
\markright{ }
\begin{abstract}
\noindent
The recent experimental measurement of the muon $g-2$ at Fermilab National Laboratory, at a $4.2\sigma$ tension with the Standard Model prediction, highlights the need for further improvements on the theoretical uncertainties associated to the hadronic sector. In the framework of the operator product expansion in the presence of a background field, the short-distance behaviour of the hadronic light-by-light contribution was recently studied. The leading term in this expansion is given by the massless quark-loop, which is numerically dominant compared to non-perturbative corrections. Here, we present the perturbative QCD correction to the massless quark-loop and estimate its size numerically. In particular, we find that for scales above 1 GeV it is relatively small, in general roughly $-10\%$ the size of the massless quark-loop. The knowledge of these short-distance constraints will in the future allow to reduce the systematic uncertainties in the Standard Model prediction of the hadronic light-by-light contribution to the $g-2$.
 
\end{abstract}
\end{frontmatter}
\section{Introduction}
The most recently updated Standard Model (SM) prediction of the muon anomalous magnetic moment, or, $a_{\mu} = (g-2)_{\mu}/2$, is~\cite{Aoyama:2020ynm}
\begin{equation}
	a_{\mu}^{\mathrm{SM}} = 116\,  591\,  810  (43)  \times 10 ^{-11}\, .
\end{equation}
Earlier this year, the Fermilab National Laboratory released its first experimental value for $a_{\mu}$~\cite{Muong-2:2021ojo}, which when combined with the previous Brookhaven National Laboratory result~\cite{Bennett:2006fi} gives
\begin{align}
	a_{\mu}^{\textrm{exp}} = 116\,  592\,  061  
	(41)   \times 10 ^{-11}\, .
	\end{align}
The difference between experiment and theory is thus
\begin{align}
	\Delta a_{\mu} = a_{\mu}^{\textrm{exp}}  - a_{\mu}^{\mathrm{SM}} = 251 (59)\times 10 ^{-11} \, ,
\end{align}
which corresponds to a $4.2\sigma$ tension. Further effort is required to improve the precision on both theoretical calculations and experimental measurements, this in order to finally understand whether this is a sign of new physics beyond the SM.

\begin{figure}[t!]\centering
	\includegraphics[width=0.2\textwidth]{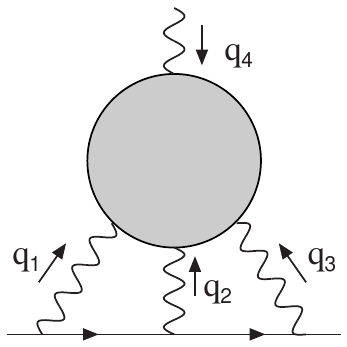}
	\caption{\label{fig:hlbltensor} The HLbL contribution $ a_{\mu}^{\mathrm{HLbL}} $, where the blob contains all possible hadrons.}
\end{figure}

On the theoretical side, the uncertainty on $a_{\mu}^{\mathrm{SM}}$ is dominated by contributions from the hadronic sector~\cite{Aoyama:2020ynm}. One of the two hadronic contributions responsible for this is the hadronic light-by-light (HLbL), $ a_{\mu}^{\mathrm{HLbL}} $. It is $ a_{\mu}^{\mathrm{HLbL}} $ that we will be concerned with here, diagrammatically represented as in Fig.~\ref{fig:hlbltensor}. This contribution can be calculated either in a dispersive data-driven approach or on the lattice~\cite{Aoyama:2020ynm,KnechtQCD21,GerardinQCD21}, and the work presented here is relevant for the former method. 

In Fig.~\ref{fig:hlbltensor} we have denoted the three virtual loop-momenta by $q_{1,2,3}$, and the external photon momentum by $q_{4}$. The external photon is soft, i.e.~in the static limit $q_{4}\rightarrow 0$. In the HLbL loop-integration several momentum scales are therefore mixed. Defining the Euclidean virtualities $Q_{i}^2 = -q_{i}^2$ for $i=1,\, 2,\, 3$ we can single out two of the kinematical regions of particular interest to us: $(a)$ The mixed region where $Q_{i}^2,\,  Q_{j}^2 \gg Q_{k}^2 , \, \Lambda _{\textrm{QCD}}^2$. Note that there is no inherent ordering between  $Q_{k}^2$ and $ \Lambda _{\textrm{QCD}}^2$. $(b)$ The purely short-distance (SD) region $Q_{i}^2 \gg \Lambda _{\textrm{QCD}}^2$. In these two regions one can perform operator product expansions (OPEs) between currents in the definition of the HLbL correlation function, giving rise to so-called SD constraints (SDCs) on the HLbL amplitude~\cite{Melnikov:2003xd,Bijnens:2019ghy,Bijnens:2020xnl,Bijnens:2021jqo}. These are relevant to constrain model calculations of some of the intermediate states arising in the dispersive treatment of the HLbL amplitude, see e.g.~Refs.~\cite{Aoyama:2020ynm,Colangelo:2019uex,Colangelo:2021nkr,Melnikov:2019xkq} for recent applications. Alternatively, they may be used to complement model calculations in the given kinematical limits. Below, we will discuss SDCs derived in the purely SD region, $(b)$ above, after a proper definition of the HLbL tensor has been given in Sec.~\ref{sec:hlbltensor}.

\section{The HLbL tensor}\label{sec:hlbltensor}
The HLbL tensor $\Pi^{\mu_{1}\mu_{2}\mu_{3}\mu_{4} } (q_{1},q_{2},q_{3})$ is defined as the correlation function of four light-quark electromagnetic currents $J^{\mu _{i}}(x_i) = \bar{q}\, Q_{q}\gamma ^{\mu}q$ for $q=(u,d,s)$ and $Q_{q}=\textrm{diag}(e_q) =\textrm{diag}(2/3,-1/3,-1/3)$ being the charge-matrix of the quarks in question. This can be written~\cite{Bijnens:2019ghy}
\begin{eqnarray}\label{eq:hlbltensor}
	\Pi^{\mu_{1}\mu_{2}\mu_{3}\mu_{4} } (q_{1},q_{2},q_{3}) = 
	-i\int \frac{d^{4}q_{4}}{(2\pi)^{4}}\left(\prod_{i=1}^{4}\int d^{4}x_{i}\, e^{-i q_{i} x_{i}}\right) 
	\hspace{-100pt}
	\nonumber \\
	\times \langle 0 | T\left(\prod_{j=1}^{4}J^{\mu_{j}}(x_{j})\right)|0\rangle \, .
\end{eqnarray}
The tensor $	\Pi^{\mu_{1}\mu_{2}\mu_{3}\mu_{4} } $ satisfies the Ward identities $q_{i,\, \mu_{i}} \, \Pi^{\mu_{1}\mu_{2}\mu_{3}\mu_{4}}=0$, for any $q_{i}$ . This in turn implies that one can rewrite the tensor in terms of its own derivative
\begin{equation}\label{eq:wardcons}
	\Pi^{\mu_{1}\mu_{2}\mu_{3}\mu_{4}}=-q_{4,\, \nu_{4}}\frac{\partial \Pi^{\mu_{1}\mu_{2}\mu_{3}\nu_{4}}}{\partial q_{4,\, \mu_{4}}} \, .
\end{equation}
From this one sees that to calculate  $a_{\mu}^{\mathrm{HLbL}}$ it is sufficient to consider the derivative above. Moreover, since the external photon is static, a Lorentz decomposition of the derivative yields only 19 independent structures. By defining 19 projectors $P^{\tilde{\Pi}_{i}}_{\mu_1 \mu_2 \mu_3 \mu_4 \nu_4}$ one can then find 19 functions
\begin{equation}\label{eq:pitildes}
	\tilde{\Pi}_{i}=P^{\tilde{\Pi}_{i}}_{\mu_1 \mu_2 \mu_3 \mu_4 \nu_4}\lim_{q_4\to 0} \frac{\partial \Pi^{\mu_{1} \mu_{2} \mu_{3} \nu_{4}}}{\partial q_{4}^{\mu_{4}}} \, .
\end{equation}
The projectors can be found in Ref.~\cite{Bijnens:2021jqo}, and are here left out for brevity. The scalar functions $\tilde{\Pi}_i$ can in turn be related to six functions $\hat{\Pi}_{1,4,7,17,39,54}$ that entirely determine $a_{\mu}^{\mathrm{HLbL}}$. The HLbL loop-integral can be written as~\cite{Aoyama:2020ynm}
\begin{eqnarray}\label{eq:amuhlblint}
	\hspace{-5pt}
	a_{\mu}^{\mathrm{HLbL}} = \frac{2\alpha ^{3}}{3\pi ^{2}} 
	\int _{0}^{\infty} dQ_{1}\int_{0}^{\infty} dQ_{2} \int _{-1}^{1}d\tau \, \sqrt{1-\tau ^{2}}\,
	\hspace{-15pt}
	\nonumber \\
	\times \,  Q_{1}^{3}Q_{2}^{3} \sum _{i=1}^{12} T_{i}(Q_{1},Q_{2},\tau)\, \overline{\Pi}_{i}(Q_{1},Q_{2},\tau)\, ,
\end{eqnarray}
where $T_{i}$ are known kernels and $\overline{\Pi}_{i}$ are linear combinations of $\hat{\Pi}_{1,4,7,17,39,54}$. The relation between the $\tilde{\Pi}_i$ and $\hat{\Pi}_j$ is straight-forward to obtain, see  Ref.~\cite{Bijnens:2021jqo}. In other words, by calculating the derivative of the HLbL tensor and projecting onto the $\tilde{\Pi}_{i}$, one can obtain the $\hat{\Pi}_{j}$ and thus $a_{\mu}^{\mathrm{HLbL}}$ can be determined from~(\ref{eq:amuhlblint}).

\begin{figure}[t!]\centering
	\includegraphics[width=0.45\textwidth]{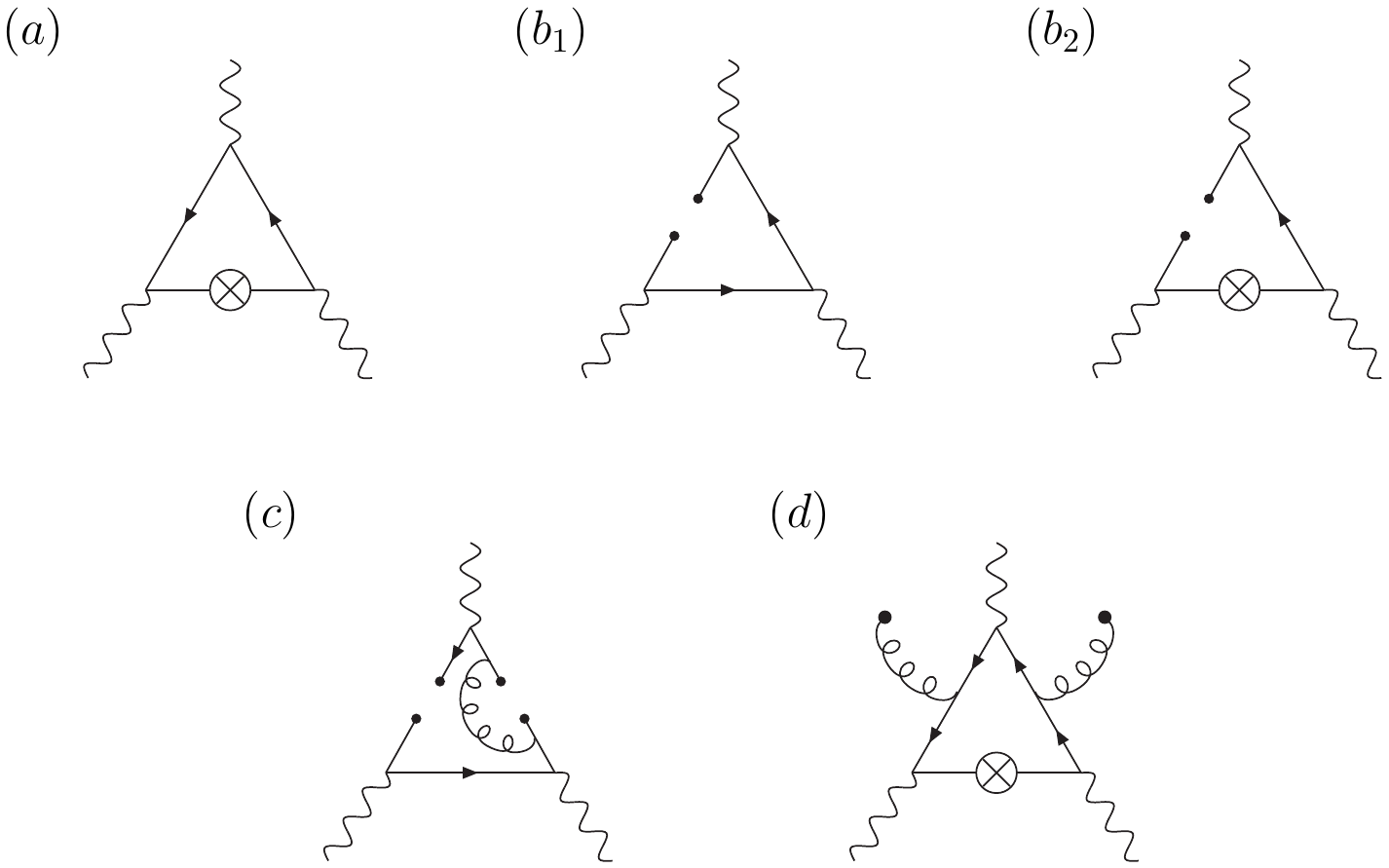}
	\caption{\label{fig:previouspaper} The previously calculated terms in the background field OPE. The insertion of the soft external field on hard lines has been indicated as a crossed vertex. In diagrams without a crossed vertex the external field induces the non-perturbative condensates present. }
\end{figure}

The integration in~(\ref{eq:amuhlblint}) is for the full HLbL contribution, but as discussed in the previous section one may divide the integration domain into different regions. In the purely SD limit where $Q_{i}^2 \gg \Lambda _{\textrm{QCD}}^2$, one can use OPE techniques to calculate the SD contribution to $a_{\mu}^{\mathrm{HLbL}}$~\cite{Bijnens:2019ghy}. The onset of this asymptotic domain is not clear, so one defines a $Q_{\textrm{min}}$ as a variable lower cut-off in the integration~(\ref{eq:amuhlblint}) via $Q_{i}^2 > Q_{\textrm{min}}^2$ which should lie above the onset. 
In Refs.~\cite{Bijnens:2019ghy,Bijnens:2020xnl} SDCs for the HLbL were derived for arbitrary $Q_{\textrm{min}}$. 
Since the external photon for the $g-2$ kinematics is soft, it is not possible to simply do an OPE of the tensor in~(\ref{eq:hlbltensor}). Instead, as was shown in Ref.~\cite{Bijnens:2019ghy}, one may treat the external photon as a static background field and consider a 3-point function in the presence of this field, i.e.~
\begin{eqnarray}
	\label{eq:3pointem}
	\Pi ^{\mu_{1} \mu_{2} \mu_{3} }(q_{1},q_{2}) = 
	-\frac{1}{e}\int\frac{d^4 q_{3}}{(2\pi)^4}  \left(\prod_{i=1}^{3}\int d^{4}x_{i}\, e^{-i q_{i} x_{i}}\right) 
	\hspace{-18pt}
	\nonumber \\
	\times  \langle 0 | T\left(\prod_{j=1}^{3}J^{\mu_{j}}(x_{j})\right) | \gamma(q_4) \rangle \, .
\end{eqnarray}
The reason this works is that $	\Pi^{\mu_{1}\mu_{2}\mu_{3}}$ can be connected to the derivative of $ \Pi^{\mu_{1}\mu_{2}\mu_{3}\mu_{4}}$~\cite{Bijnens:2019ghy}. The leading term in this OPE is the massless perturbative quark-loop (see Fig.~\ref{fig:previouspaper}(a)) which scales as $1/Q^2$~\cite{Bijnens:2019ghy}. Massive corrections to this as well as leading and next-to-leading non-perturbative corrections (see Fig.~\ref{fig:previouspaper}(b)--(d)), scaling up to $1/Q^6$, were calculated and numerically estimated in Refs.~\cite{Bijnens:2019ghy,Bijnens:2020xnl}. For $Q_{\textrm{min}}\gtrsim 1$ GeV, it was found that these corrections were numerically suppressed with respect to the massless perturbative quark-loop by at least one to two orders of magnitude, as reported in this conference last year~\cite{Bijnens:2020yfc}. This lies beyond the current precision goal on the HLbL. Note that 2-loop gluonic perturbative corrections to the massless quark-loop still can be sizeable. Below, we present our calculation of these contributions. Further details can be found in Ref.~\cite{Bijnens:2021jqo}.

\section{2-loop calculation}
\begin{figure}[t!]\centering
	\includegraphics[width=0.4\textwidth]{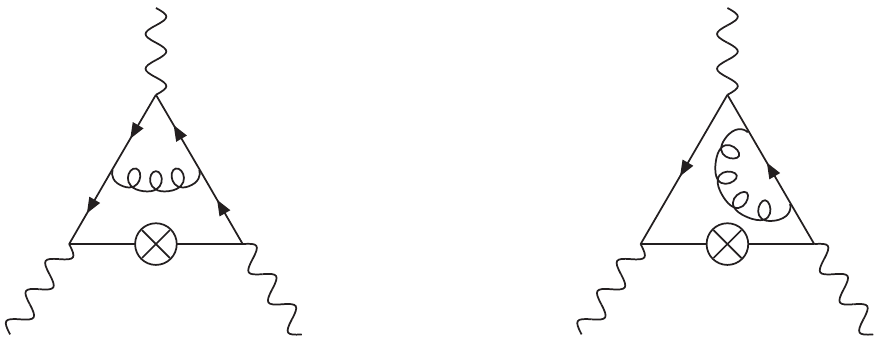}
	\caption{\label{fig:2looptops} Examples of the 2-loop topologies. The insertion of the soft external field on the hard quark-lines has been indicated as a crossed vertex.}
\end{figure}

\begin{figure}[t!]\centering
	\includegraphics[width=0.5\textwidth]{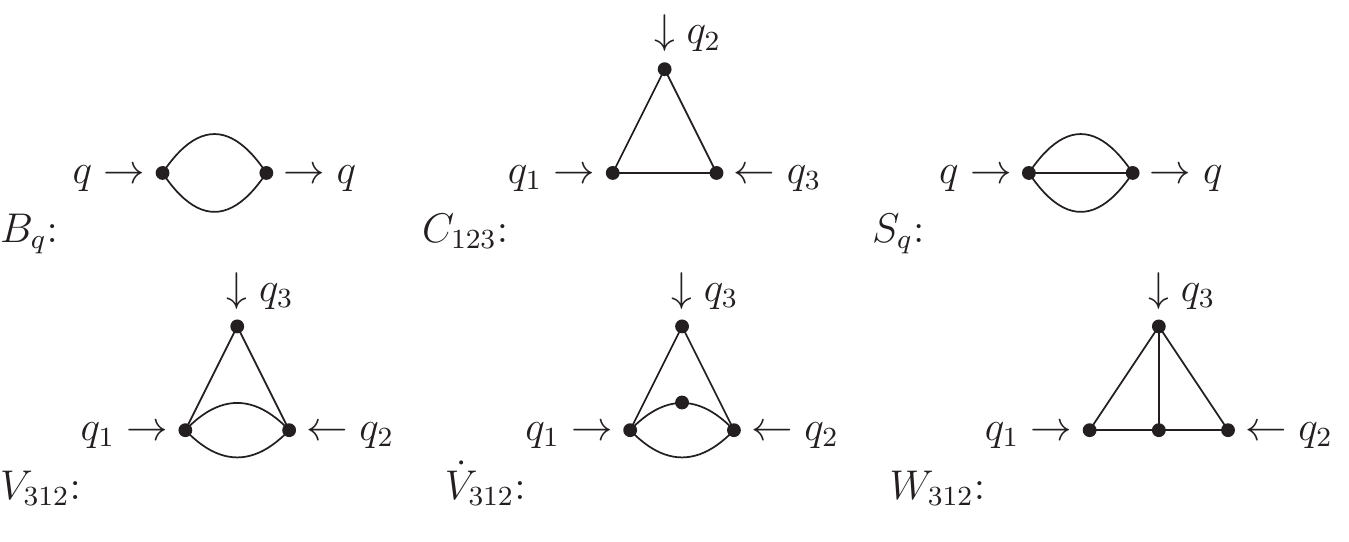}
	\caption{\label{fig:masterints} The six different master integrals contributing.}
\end{figure}

Diagrammatic examples of the 2-loop gluonic corrections are shown in Fig.~\ref{fig:2looptops}. These are given from two insertions from the Dyson series into~(\ref{eq:hlbltensor}), i.e.~
\begin{align}
	\label{eq:backdysonqcd}
	\Pi ^{\mu_{1} \mu_{2} \mu_{3} \mu_{4}}
	&
	=
	-i\int\frac{d^4 q_{3}}{(2\pi)^4}  \left(\prod_{i=1}^{4}\int d^{4}x_{i}\, e^{-i q_{i} x_{i}}\right) 
	\nonumber \\
	&
	\times 	\langle 0 | T\Bigg\{                                                \prod_{j=1}^{4}J^{\mu_{j}}(x_{j})
\,  \frac{1}{2}\prod _{i=1}^2  \int d^4 z_{i} \,
i \mathcal{L}_{\textrm{int}}^{\textrm{qgq}} (z_{i}) 
	\Bigg\} |0\rangle \, ,
\end{align}
where 
\begin{align}
\mathcal{L}_{\textrm{int}}^{\textrm{qgq}}(z_{i}) = g_{S}\frac{\lambda ^{a_{i}}_{\bar{\gamma }_{i}\bar{\delta}_{i}}}{2}
B^{a_{i}}_{\nu _{i}}(z_{i}) \, \bar{q}^{\bar{\gamma} _i}(z_i)\gamma ^{\nu _i}q^{\bar{\delta} _i}(z_i) \, ,
\end{align}
for the strong coupling $g_S$, gluon field $B^{a_{i}}_{\nu _{i}}$ and Gell-Mann matrices $\lambda ^{a_{i}}$ with barred colour-space indices.
Taking the derivative of~(\ref{eq:backdysonqcd}) with respect to $q_{4}$ and the static limit, one can project onto the $\tilde{\Pi}_{i}$ as in~(\ref{eq:pitildes}). The $\tilde{\Pi}_{i}$ are then given as linear combinations of roughly 6000 scalar 2-loop integrals with three external scales $q_{1,2,3}^2$, defined in $d=4-2\epsilon$ dimensions. 
To reduce this large set of integrals we use {\sc KIRA}~\cite{Maierhoefer:2017hyi}, and a minimal set of six master-integrals to describe them is shown diagrammatically in Fig.~\ref{fig:masterints}. The exact definitions of these are given in Ref.~\cite{Bijnens:2021jqo}. Note that although the $\tilde{\Pi}_i$ are all separately finite without need of renormalisation, the six master-integrals are separately divergent in $\epsilon$. These divergences have to cancel. The needed expansions in $\epsilon$ are provided in Refs.~\cite{Birthwright:2004kk,Chavez:2012kn}\footnote{As explained in Ref.~\cite{Bijnens:2021jqo}, we found minor typos in the expansions in Ref.~\cite{Chavez:2012kn}. }. The finite expressions for the $\tilde{\Pi}_{i}$ are of the form
\begin{align}\nonumber
	\tilde{\Pi}_{m}&=f_{m,ijk}^{pqr}F_{ijk}(2)Q_1^{2p}Q_2^{2q}Q_3^{2r}+w_{m,ijk}^{pqr}W_{ijk}(0)Q_1^{2p}Q_2^{2q}Q_3^{2r}
	\nonumber \\
	&
	+c_{m,ijk}^{pqr}C_{ijk}(0)Q_1^{2p}Q_2^{2q}Q_3^{2r}
+n_{m,1}^{pqr}Q_1^{2p}Q_2^{2q}Q_3^{2r}\log\frac{Q_{1}^2}{Q_{3}^2}
\nonumber \\
&
+n_{m,2}^{pqr}Q_1^{2p}Q_2^{2q}Q_3^{2r}C_{ijk}(0)\log\frac{Q_{2}^2}{Q_{3}^2} \nonumber
\nonumber \\
&
+l_{m,ijk1}^{pqr}Q_1^{2p}Q_2^{2q}Q_3^{2r}C_{ijk}(0)\log\frac{Q_{1}^2}{Q_{3}^2}
	\nonumber \\
	&
	+l_{m,ijk2}Q_1^{2p}Q_2^{2q}Q_3^{2r}C_{ijk}(0)\log\frac{Q_{2}^2}{Q_{3}^2}\, ,
\end{align}
where $F_{ijk}(2)$, $W_{ijk}(0)$ and $C_{ijk}(0)$ are finite coefficients from the $\epsilon$--expansions. These can used in linear combinations to yield the $\tilde{\Pi}_{j}$ needed for $a_{\mu}^{\mathrm{HLbL}}$. In the supplementary material of Ref.~\cite{Bijnens:2021jqo} we provide all of these analytical results. The numerical evaluation of $a_{\mu}^{\mathrm{HLbL}}$ is discussed in the next subsection.

\section{Numerical evaluation}
\begin{figure}[t!]\centering
	\includegraphics[width=0.35\textwidth]{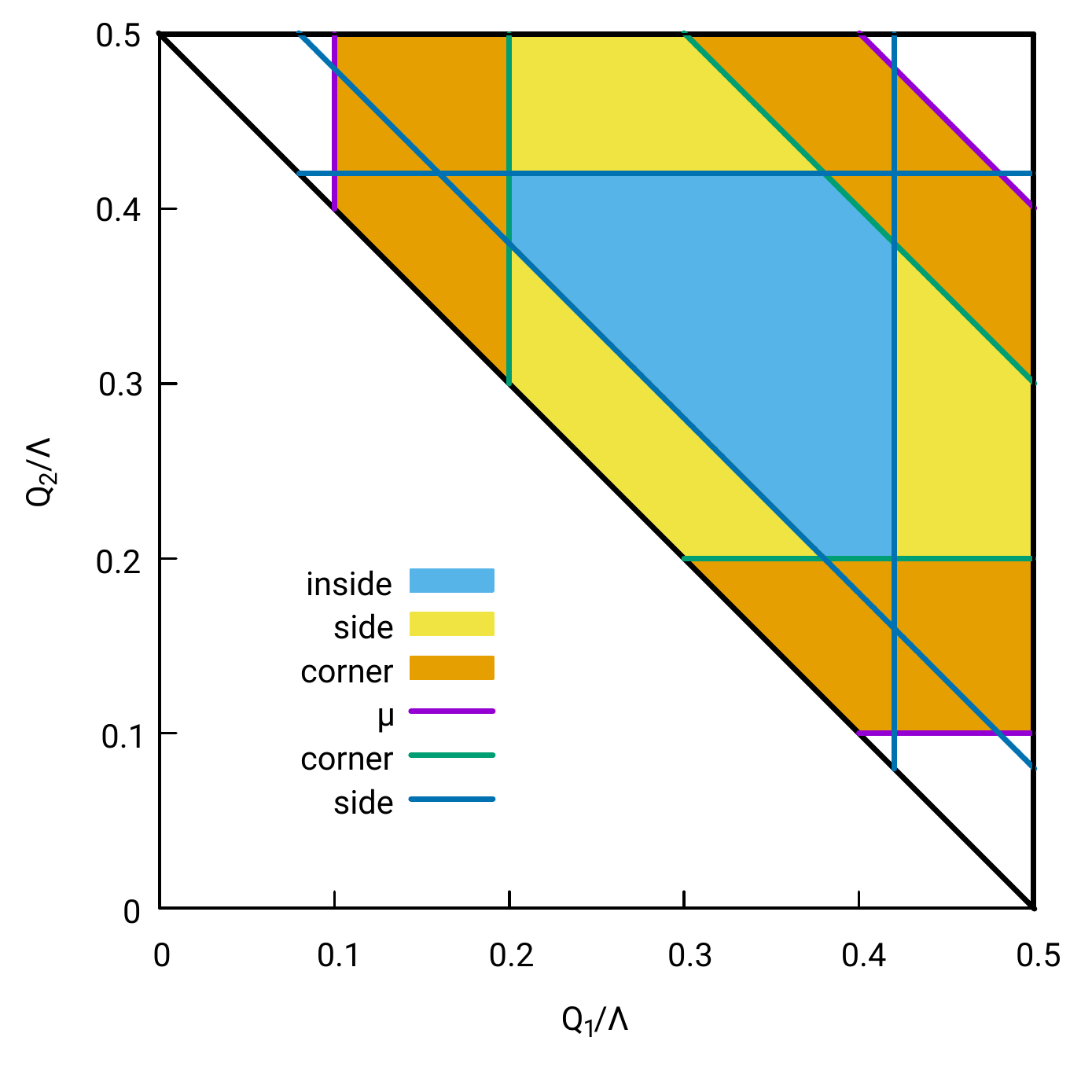}
	\caption{\label{fig:triangle} The integration domain of $a_{\mu}^{\textrm{HLbL}}$ divided into different regions. }
\end{figure}

Combining the $\tilde{\Pi}_{i}$ into $\hat{\Pi}_{j}$ analytically is very simple, but unfortunately gives rise to numerical problems in the loop-integration~(\ref{eq:amuhlblint}). The reason is the appearance of negative powers of the K\"all\'{e}n function $\lambda= (Q_1^2+Q_2^2-Q_3^2)^2-4Q_1^2 Q_2^2$ in the $\hat{\Pi}_{j}$, which at certain points in the integration domain are close to zero. These spurious divergences of course have to cancel in the end with zeros in numerators. This corresponds to points where $Q_{i} \approx Q_{j}+Q_{k}$ for some $i$, $j$ and $k$, a region in the integration domain which we denote the side region. Also the so-called corner regions where $Q_{i}\ll Q_{j},Q_k $ at the same time as $Q_{j}\approx Q_k$ prove to be problematic. We can thus divide the integration domain in the $(Q_{1},Q_2) $--plane as in Fig.~\ref{fig:triangle}, for a fixed $\Lambda = Q_1+Q_2+Q_3$ and $\mu = Q_{\min}$. We have performed analytic Taylor expansions in the small parameters for each of the problematic side and corner regions, which when implemented into a numerical integration routine give a finite $a_{\mu}^{\mathrm{HLbL}}$. These analytical expressions are provided in the supplementary material of Ref.~\cite{Bijnens:2021jqo}. We note in the passing that the corner expansions in fact correspond to a subset of the mixed region, namely $Q_{i}^2,\,  Q_{j}^2 \gg Q_{k}^2 \gg \Lambda _{\textrm{QCD}}^2$. Extending the work in Refs.~\cite{Melnikov:2003xd,Colangelo:2019uex} for $Q_{i}^2,\,  Q_{j}^2 \gg Q_{k}^2 , \, \Lambda _{\textrm{QCD}}^2$ by going to higher orders should therefore reproduce our results by choosing $Q_{k}^2\gg  \Lambda _{\textrm{QCD}}^2$.

Having taken care of the numerical issues described above we can perform the loop-integration in~(\ref{eq:amuhlblint}). First of all,  in Table~\ref{tab:intQ1} we separately present for $Q_{\textrm{min}} = 1$ GeV the values of the twelve contributions to $a_{\mu}^{\textrm{HLbL}}$ from the sum in~(\ref{eq:amuhlblint}), both for the massless quark-loop and the gluonic corrections themselves. As can be seen, the latter are, with $\alpha _s \approx 0.33$ at $1$ GeV, roughly $-10\%$ of the massless quark-loop. 
\begin{table}[t!]\centering
	\begin{tabular}{|l|r|r|}
		\hline        & Quark-loop & Gluon-loop $\times \left( \frac{\alpha_{s}}{\pi}\right) ^{-1}$ \\ \hline
		$\bar{\Pi}_{1}$  & $0.0084$   & $-0.0077$                                                                          \\ \hline
		$\bar{\Pi}_{2}$  & $13.28$  & $-12.30$                                                                       \\ \hline
		$\bar{\Pi}_{3}$  & $0.78$     & $-0.87$                                                                          \\ \hline
		$\bar{\Pi}_{4}$  & $-2.25$    & $0.62$                                                                 \\ \hline
		$\bar{\Pi}_{5}$  & $0.00$     & $0.20$                                                     \\ \hline
		$\bar{\Pi}_{6}$  & $2.34$     & $-1.43$                                                                           \\ \hline
		$\bar{\Pi}_{7}$  & $-0.097$   & $0.056$                                                                            \\ \hline
		$\bar{\Pi}_{8}$  & $0.035$    & $0.41$                                                                         \\ \hline
		$\bar{\Pi}_{9}$  & $0.623$    & $-0.87$                                                                           \\ \hline
		$\bar{\Pi}_{10}$ & $1.72$     & $-1.61$                                                                             \\ \hline
		$\bar{\Pi}_{11}$ & $0.696$   & $-1.04$                                                                            \\ \hline
		$\bar{\Pi}_{12}$ & $0.165$    & $-0.16$                                                                            \\ \hline
		Total            & $17.3$     & $-17.0$                                                                      \\ \hline
	\end{tabular}
	\caption{The twelve contributions to $a_{\mu}^{\textrm{HLbL}}$ from the sum in~(\ref{eq:amuhlblint}) with $Q_{\mathrm{min}}=1\, \mathrm{GeV}$. Units are in $10^{-11}$.}
	\label{tab:intQ1}
\end{table} 

\begin{figure}[t!]\centering
	\includegraphics[width=0.52\textwidth]{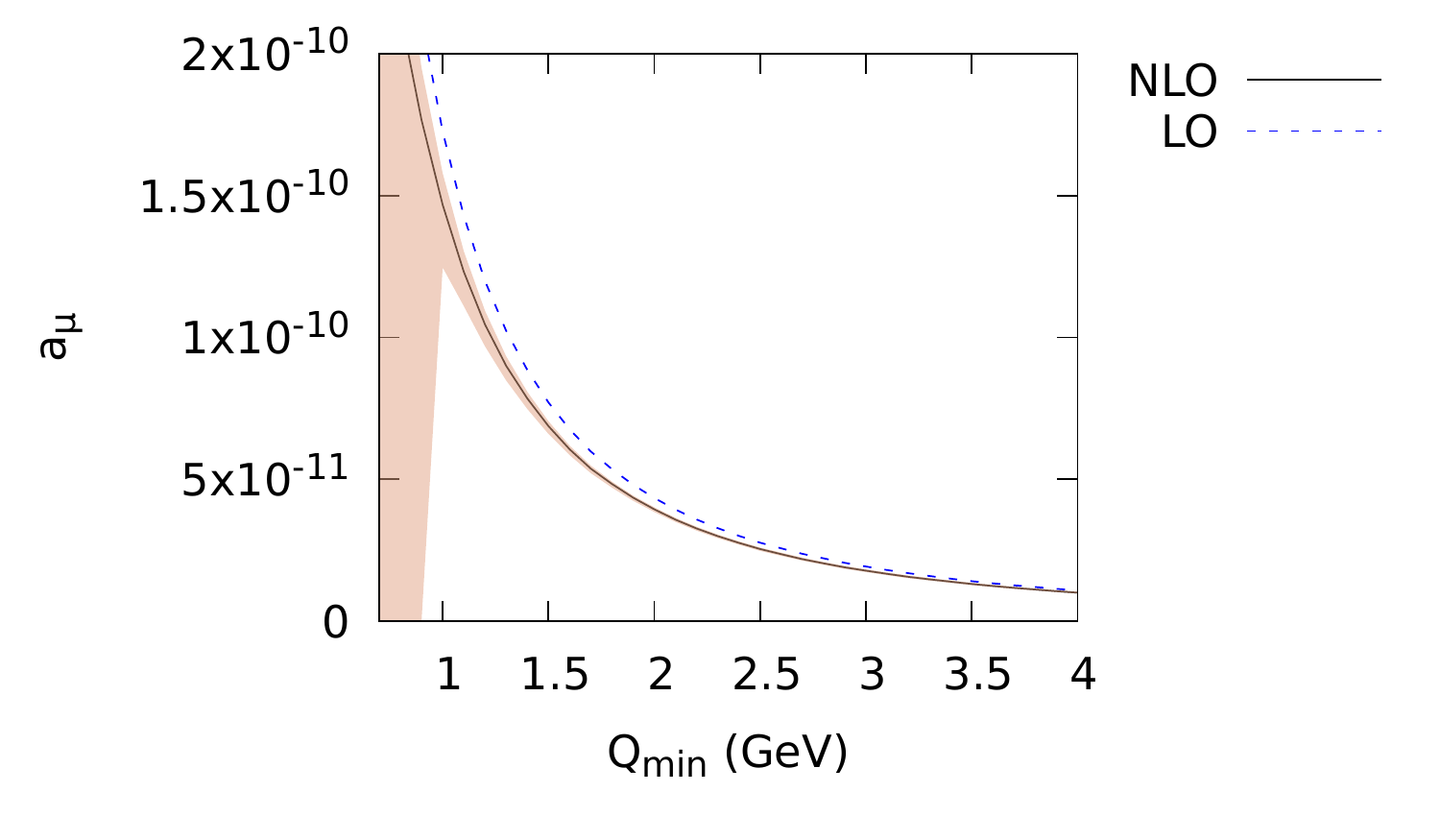}
	\caption{\label{fig:numnlo} Contributions to $a_{\mu}^{\textrm{HLbL}}$ from the massless perturbative quark-loop (LO), as well as the gluonic correction calculated herein (NLO). }
\end{figure}

In Fig.~\ref{fig:numnlo} we plot the massless quark-loop (LO) as well as the massless quark-loop together with the gluonic corrections (NLO) as functions of $Q_{\textrm{min}}$. The error band on the NLO correction comes from the choice of scale $\mu$ where $\alpha _s(\mu)$ is defined. In order to estimate this uncertainty, we first of all vary $\mu^2\in (1/2,2) Q_{\textrm{min}}^2$. In addition, we take into the uncertainty the running of $\alpha_{s}^{N_{f}=5}(M_{Z})$, by running it at five loops to both $\alpha_{s}^{N_{f}=3}(m_{\tau})$ and $\alpha_{s}^{N_{f}=3}(Q_{\textrm{min}})$. From this we clearly see the perturbative breakdown of QCD at low energies below $1$ GeV. Again, we see the general  $-10\%$ correction from the gluon corrections above $1$ GeV.

\section{Conclusions}
We have in a series of papers derived SDCs for the HLbL tensor in the purely SD region where $Q_{i}^2 \gg \Lambda _{\textrm{QCD}}^2$ are all large~\cite{Bijnens:2019ghy,Bijnens:2020xnl,Bijnens:2021jqo}. We have shown that the leading term in the background field OPE used to derive these SDCs is the massless perturbative quark-loop. With non-perturbative and massive corrections to the leading term being numerically insignificant for the current precision goal on $a_{\mu}^{\textrm{HLbL}}$, only the 2-loop gluonic corrections to the massless quark-loop are relevant, which in general are $-10\%$ of the size of the leading term. In the future, we will use OPE techniques to study SDCs for the mixed region $Q_{i}^2,\,  Q_{j}^2 \gg Q_{k}^2 , \, \Lambda _{\textrm{QCD}}^2$, which will go beyond the current results in Refs.~\cite{Melnikov:2003xd,Colangelo:2019uex}. 
 
 \section*{Acknowledgements}
 N.~H.--T. and L.~L.~are funded by the Albert Einstein Center for Fundamental Physics at Universit\"{a}t Bern and the Swiss National Science Foundation respectively. J.~B.~is supported in part by the Swedish Research Council grants contract numbers 2016-05996 and 2019-03779. A.~R.--S.~is partially supported by the Agence Nationale de la Recherche (ANR) under grant ANR-19-CE31-0012 (project MORA).


\bibliographystyle{elsarticle-num}
\bibliography{refsqcd21}

\end{document}